\documentclass[aps,pra,twocolumn,nofootinbib]{revtex4-1}

\usepackage{amsfonts}
\usepackage{amsmath}
\usepackage{graphicx}
\usepackage{amssymb}
\usepackage{times}
\usepackage{color}

\def\>{\rangle}
\def\<{\langle}

\newcounter{myenumi}
\newenvironment{myenumerate}{\begin{enumerate}\itemsep-0.3em \setcounter{enumi}{\themyenumi}}{ \setcounter{myenumi}{\theenumi}\end{enumerate}}

\begin{document}

\title{Why I am optimistic about the silicon-photonic route to quantum computing}
\author{Terry Rudolph}
\affiliation{Department of Physics, Imperial College London, London SW7 2AZ, United Kingdom}
\date{\today}

\begin{abstract}
This is a short overview\footnote{This article is the extended text of a talk I planned to give a couple of places in the United States this year, but will not do so now having been denied a visa (apparently no immigration officer likes to hear the words ``Iran'' and ``physics'' in the same sentence).} explaining how building a large-scale, silicon-photonic quantum computer has been reduced to the creation of good sources of 3-photon entangled states (and may simplify further). Given such sources, each photon need pass through a small, constant, number of components, interfering with at most 2 other spatially nearby photons, and current photonics engineering has already demonstrated the manufacture of thousands of components on two-dimensional semiconductor chips with performance that allows the creation of tens of thousands of photons entangled in a state universal for quantum computation. 

At present the fully-integrated, silicon-photonic architecture we envisage involves creating the required entangled states by starting with single-photons produced non-determistically by pumping silicon waveguides (or cavities) combined with on-chip filters and nanowire superconducting detectors to herald that a photon has been produced. These sources are multiplexed into being near-deterministic, and the single photons then passed through an interferometer to non-deterministically produce small entangled states - necessarily multiplexed to near-determinism again. This is followed by a `ballistic' scattering  of the small-scale entangled photons through an interferometer such that some photons are detected, leaving the remainder  in a large-scale entangled state which is provably universal for quantum computing implemented by single-photon measurements. 

There are a large number of questions regarding the optimum ways to make and use the final cluster state, dealing with static imperfections, constructing the initial entangled photon sources and so on, that need to be investigated before we can aim for millions of qubits capable of billions of computational time-steps. The focus in this article is on the theoretical side of such questions.

\end{abstract}
\maketitle

\section{Introduction}

Photonics is the `ugly duckling' of approaches to quantum computing. There are two overarching reasons to try and nurture it into a swan. 

The first reason is that in photonic integrated circuits (PICs) we believe that we can reduce stochastic noise levels several orders of magnitude below even optimistic estimates of such noise for matter-based approaches. For instance, raw stochastic error rates of $10^{-6}$ would, I believe, be a pessimistic estimate. (See Section\ref{sec:switches} for an argument why current experiments indicate they are no bigger than $10^{-5}$ and why I would expect at least $10^{-8}$ is achievable with current devices). However, in photonics we have an inescapable (constant) overhead, due to the fact we use non-deterministic linear optical gates, of around 100 physical photons to end up with a single ``raw computational'' photon that is the equivalent of one normal qubit. Are we are crazy? From this perspective no one has ever built a single `functional' photonic qubit yet! The reason we should not be committed to an asylum (yet) is that the overhead photonics assumes will be subsumed by the standard error correction required. For example, in \cite{fowler:12}, the authors calculate it requires $2\times 10^5$ matter qubits with a raw error rate of $10^{-3}$ (as well as many thousands of gates and measurements etc) to achieve a magic state \cite{Bravyi2005} with stochastic error rate of $3\times10^{-6}$. 

The second reason is that PICs are being vigorously pursued for classical computing purposes, and the core components necessary for the quantum architecture are already under investigation and optimization, in multiple variations, by classical photonics engineers. Moreover, the vast majority of this work is fully compatible with present-day fabrication techniques and standards (i.e. CMOS technology), so that while current classical PICs only contain several thousands of components (in a few hundred square microns) \cite{sun2013large}, it is reasonable to expect PICs containing many hundreds of thousands of components soon. This has become even more likely with recent breakthroughs \cite{Gentry:15,Sun:2015} that prove PICs can not only be CMOS compatible, they can be built with `zero-change' of current CMOS fabrication facilities. As such, contemplating PICs of many millions of photonic components is brave, but not unreasonable.
One may wonder if millions of components are really necessary? Earl Campbell and Joe O'Gorman have recently improved magic-state type protocols (which are some of the most resistant to stochastic noise) and have done careful estimates of resources required for running Shor's algorithm to factorize numbers beyond those classically achievable. Their numbers indicate we need to be able to swallow resource numbers \emph{qubits}$\times$\emph{time steps} $\approx 10^{15}\mathrm{-}10^{17}$. CMOS technology is the only method known for manufacturing dynamical devices on this scale.
 
Typically matter-based approaches to quantum computing begin by building and controlling small numbers of qubits; the challenge is then scaling up. For the photonic architecture I will overview here, things are somewhat inverted: the primary challenge is efficiently producing small (3-photon) entangled states on a PIC. The large scale architecture becomes relatively trivial - it is just an interferometer built from components that are already able to be fabricated to high precision, sufficient to generate large amounts of entanglement in a cluster state \cite{Raussendorf2001} universal for quantum computing. Given such sources our overhead will then be less than 20 physical photons per final qubit in the cluster state, only two photons would undergo any kind of potentially noisy active element, no proper quantum memory is required, and all photons go through a constant number of photonic elements regardless of the computation size.  Because the architecture is based on a highly modular production of cluster state, once we do have a single functional photonic qubit we not only have arbitrary numbers of such qubits, we also have the ability to perform arbitrary gates between them. 

None of the preceding should be taken as advocating the abandonment of matter-based approaches (see Appendix A for a rant about this). Rather I want here to emphasize the considerable amount of `photonics specific' theory that remains to be done. Matter-based approaches benefit from many theoretical results that can abstract away the underlying physics of any particular implementation. However, most of this work is of marginal relevance to the key challenges of building a photonic quantum computer.

\section{Optical frequency photons as qubits}\label{sec:photonsasqubits}

For most physical realizations of a qubit, the natural environment in which it is immersed is a source of stochastic noise. By stochastic noise I will mean any random evolution of the qubit, the unknown specific realization of which differs from one computational time step to another. For some matter-based realizations it is now possible to keep the qubit fairly well isolated until the time a gate needs to be performed on it, and the majority of stochastic errors arise when it is manipulated. Almost all theory of quantum error correction and fault tolerance is based around combating the most common such errors. 

For optical frequency photons manipulated by passive interferometers built from essentially perfect linear dielectric  - such as waveguides etched into a semiconductor chip \cite{Politi2008} - environmental stochastic noise is presently unmeasurable, and expected to be very small. Essentially the energy scales (optical for the photon versus microwave for even room temperature surroundings) are too disparate for this to form any kind of worry until very large scale photonic devices are in hand. I am fond of glibly pointing out that photonic quantum information has already been proven stable against thermal decoherence for 11.5 billion years, which is the time light emitted from the Lyman-$\alpha$ blob LAB-1 retains its polarization on its journey to earth. This is close to the theoretically maximum achievable lifetime for any qubit of 13.7 billion years.


This is not to say there will be no stochastic noise on photonic qubits. The most relevant sources of such noise are the two \emph{active} elements of the architecture, which are adjustable phase shifters and switches. Essentially they are a source of stochastic noise because we cannot expect them to operate identically from one use to the next. The imperfection is primarily that of the imprecision of the voltage control. In Section \ref{sec:switches} I will discuss switches and phase shifters in more detail. The over-arching challenge of the architecture discussed here is to minimize the use of active components.


\subsection{Photons need to be indistinguishable from their neighbors}

In addition to switches and phase shifters, our photon sources need considerable improvement. They are discussed in detail in Section~\ref{sec:sources}, but at a high level it is worth noting that the architecture we are proposing requires every photon to interfere with at most 3 other photons, spatially nearby, on the PIC. For the interference to work, ideally the photons would have the same wavepacket \cite{rohde2006error}. It is not important per se whether they are a single frequency, only that their wavepackets are the same, but one way to ensure they are close to identical is to filter the photons as close to a single frequency mode as possible. For this we can use integrated ring resonators \cite{Shen:11b,filterandsource}. Filtering increases purity at the expense of increasing loss, but it is better to have loss, which is directly detectable, than undetectable stochastic errors. Impurity of single-mode entangled states can be distilled out extremely efficiently, limited only by the negligible thermal stochastic noise of the interferometers \cite{panpurification}.  

To enhance interference it is also possible to locally ``trim'' the components of the PIC \cite{Shen:11b, Schrauwen:08,Bachman:13,MarshallPC} using lasers, and this allows extremely precise manipulation of the final single photon state. (Trimming is also used extensively on classical integrated circuits.) Such trimming is `set and forget', i.e once it is done it doesn't require maintenance or consume power, it changes the static properties of the device. 

\subsection{Static imperfection is not stochastic noise}

Manufacturing imperfections on PICs are inescapable, but they are static, they do not drift with time. If the details of static imperfection (into which we can subsume other static error like imperfections of the photon wavepacket, inaccuracy of the voltage controls etc) are unknown then this leads to systematic error. However photonics is unique (I think) in that we can use a classical field (laser light) to determine the interferometer's parameters (the mode transformation it effects) to very high accuracy. These are the same parameters as those that govern the 2-photon interference crucial to the whole computer. Such characterization can also be achieved with repeated use of single photons instead, if such sources are more readily at hand on the chip. So, one way or the other, we actually expect to have very well known systematic imperfection to deal with. 


Unknown systematic error is typically more benign than stochastic noise, and can (for example) be dealt with algorithmically \cite{Reichardt2005}, amongst other ways. Known systematic imperfection is much more benign again, although we have done little theory to really explore the implications. Some general things can be said however. Computational universality in the measurement-based setting we envision does not require a cluster state. Having a known state `close to' the cluster state will be sufficient, and rather than trying to somehow correct the state and then use it we likely would do better to simply use the ideas of \cite{Gross2007} (and generalizations thereof) to perform computation with it directly in a way that effectively corrects the imperfection via the side classical computation that chooses measurement bases. If we are really lucky (a vague argument has been made to me that we are, I'm a little skeptical) then systematic imperfection produces a state within the known (symmetry-protected) `phase' of states in the same universality class as the cluster state \cite{Else2012}. If so then even potentially complicated algorithmic correction will be unnecessary.


When I say a known state `close to' the cluster state suffices, this does not mean we need produce a state with near unit fidelity (overlap with) the ideal cluster state. In fact it could have vanishing overlap with it, because local imperfections can multiply up to drive the state far from the ideal one, and still work fine. For example, while we don't yet have a complete picture of exactly how interferometer imperfections translate into logical imperfection of the large scale cluster state discussed in Section \ref{sec:logicalarchitecture}, in simple examples the final $N$-qubit state takes the form 
\[
\prod_{i\sim j} \widetilde{CZ_{ij}}|\tilde{+}\>_i^{\otimes N}
\]
where $\widetilde{CZ_{ij}}$ and $|\tilde{+}\>$ are `close to' the ideal controlled-$Z$ between qubits $i$ and $j$ and +1 eigenstate of Pauli $X$. Consider the simple case that $\widetilde{CZ_{ij}}=CZ_{ij}$, while $|\tilde{+}\>_i$ varies (but is known well) over each qubit. Then there are simple and highly precise filters (beamsplitter followed by detection), which we can apply to each photon individually, that with success probability dependent on $|\langle +|\tilde{+}\>_i|^2$, either turn this into the ideal cluster state \cite{Barrett2009}, or remove the qubit in the standard way (i.e. via an effective computational basis measurement). 

The reason I emphasize this over-simple example is that it demonstrates that although there may be a finite  systematic imperfection on every qubit which varies over the whole device, the state still has infinite localizable entanglement (and is computationally universal) as long as $|\langle +|\tilde{+}\>_i|^2$ is close enough to 1. (How close depends on details of the cluster state created. For the bond-percolated lattice discussed in the next section $|\langle +|\tilde{+}\>_i|^2\sim 0.96$ would suffice.)  To reiterate, the final state might have vanishing overlap with the ideal state - we just need to know what the actual state is as precisely as possible. Ideally we would rather avoid a two-stage process of `state correction' followed by computation, and rather just compute directly on the imperfect state.  


 Here are some of the theoretical questions which remain largely unexplored on this topic:
\begin{myenumerate} 
\item Is there a simple model that captures how do manufacturing imperfections translate into logical imperfection of the large scale entangled state we produce? If so, can we apply the wealth of existing abstract theory regarding states universal for MBQC to determine how best to use these states? 
\end{myenumerate}

\section{Logical overview of the architecture}\label{sec:logicalarchitecture}

\begin{figure}
\centering
\includegraphics[width=\linewidth]{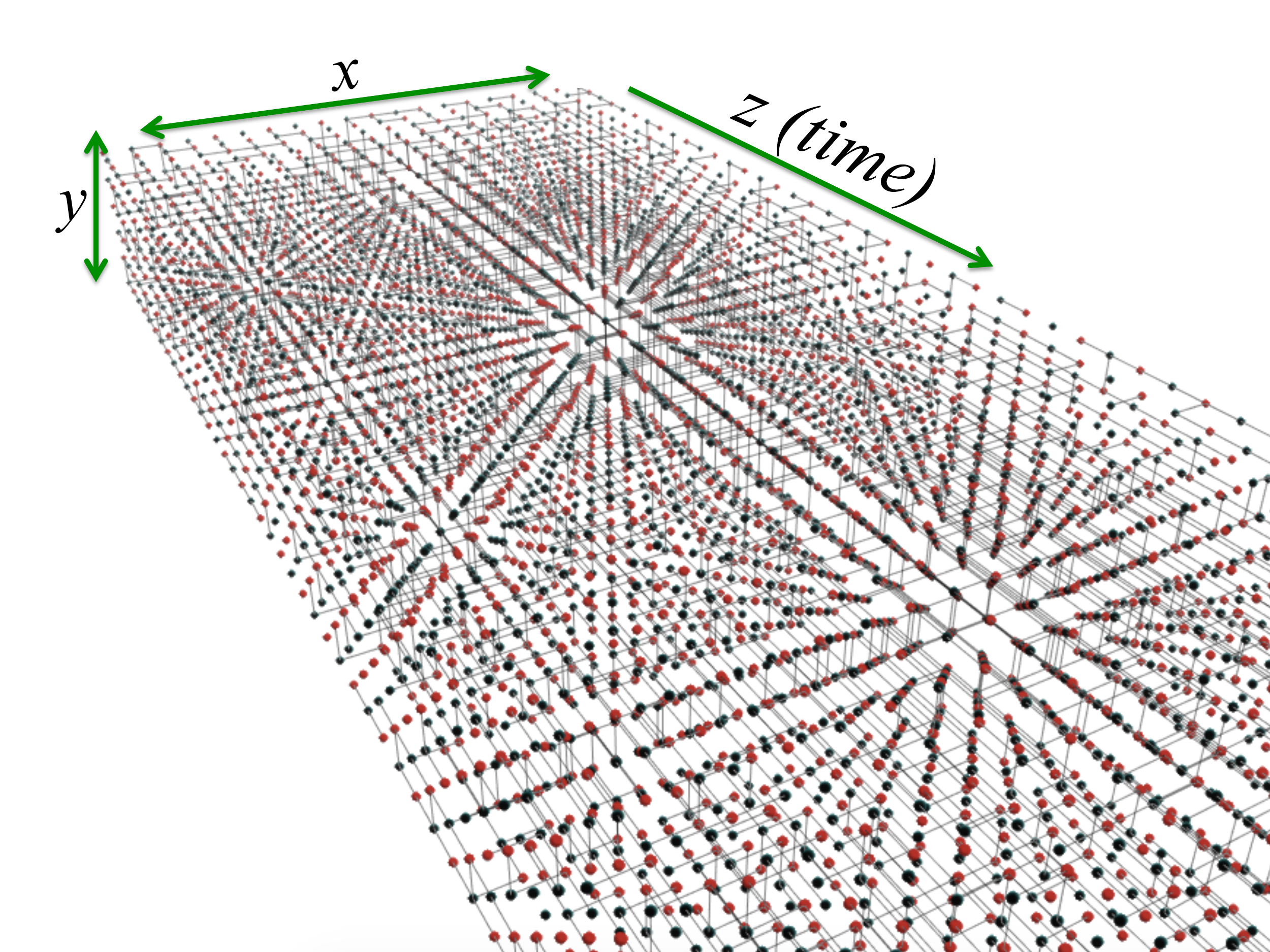}
\caption{An example of a bond-percolated surface code cluster state that can be generated on a 2d photonic integrated circuit. Only a constant number of layers of the cluster in the $z$-direction will need to be present at any particular time. The red/black coloring of qubits is primarily to aid the eye, though for this lattice correspond to primal/dual qubits of the scheme in \cite{Raussendorf2006,raussendorfharringtongoyal}. Even for only a very small thickness - here 6 qubits - in the $y$-direction the percolated lattice will extend $O(10^4)$ qubits in the $y$ and $z$ directions.}
\label{fig:lattice}
\end{figure}

At the logical level, the current architecture we are proposing creates a large entangled cluster state \cite{Raussendorf2001}, with several layers of the cluster ``alive'' at the same time. The algorithm implemented is determined by the choice of phase shifts (as computed by an efficient classical side-computation) that some of the photons experience before measurement, all the rest of the architecture is fully modular. At this logical level the cluster state being produced is a bond-percolated 3d lattice, as depicted in Fig.~\ref{fig:lattice}. Lattices where each vertex only has degree 4, such as diamond lattice or the``surface code'' lattice of \cite{Raussendorf2006}, seem best; although they are not the only ones achievable from the 3-photon GHZ states which will be our fundamental resource \cite{Zaidi2015}.

In Fig.~\ref{fig:lattice}, the $z$-dimension of the cluster state shown should be thought of as time. Thankfully, not all the photons depicted would need to be ``alive'' simultaneously; in fact, in cluster state computation, in principle one need only to have the qubits within two bonds of the qubit currently being measured actually alive. However, due to the percolated nature of the lattice, the choice of measurements to make on the leading layer of photons depends on features of the lattice behind that layer. There is therefore a certain `computational window' of qubits which will necessarily need to be kept alive within the computer while choices of measurement basis are made.

There are two ways we can use such a lattice in a computation. In Method 1, which was the original proposal \cite{Kieling2007}, we simply carve through paths that encode single qubits as per standard cluster state computation. Preliminary simulations show that we can keep such paths going essentially indefinitely with a computational window of about 10-15 (this means we need a delay line of about 10-15 clock cycles, see Section \ref{sec:delays}). We also find that, although we haven't yet incorporated any coding against loss, there is actually some serendipitous loss tolerance, on the order of a few percent. Specifically, detecting a lost photon we can attempt to logically punch it out using computational basis measurements. This works because our lattice has percolated quite far above the percolation threshold. Note that this recovery method is almost certainly not optimal even for Method 1 - various stabilizers can be measured to remove lost qubits as well. 

Method 2 is to actually use the surface code from which this lattice originates, as per the ideas in \cite{raussendorfharringtongoyal}. Given that this method automatically provides some tolerance to stochastic noise, it is likely preferable. However, unlike the lattice assumed there, here we are missing bonds (the specific locations of which we know in advance of performing the computational measurement). We also will have some photon loss that we only know has occurred at the point of detection. The computational method of \cite{raussendorfharringtongoyal} involves carving out deformable `tubes' measured in the computational basis (essentially the equivalent of the single qubit lines in Method 1) with in-between (`vacuum') qubits measured in the $|\pm\>$ eigenbasis of Pauli-$X$. Both the missing bonds and loss can potentially be tackled via the method of `super-stabilizers', where we combine (in a side classical computation) stabilizers into larger ones, and deform the tube surfaces suitably to swallow errors into tube interiors. This technique makes regular surface code computation highly resilient to loss \cite{Stace2009,Barrett2010}, despite the fact it is, of course, a code designed to deal with stochastic error. So our focus at present is trying to understand if Method 2 really will be better than Method 1 (a combined option in which we first `repair' the surface code lattice similar to Method 1 also exists), but a full simulation and comparison has not yet been done.     

Note that once we have built the cluster state, there is no difference between performing a logical two-qubit gate versus a one-qubit gate. That difference is one only of the pattern of single photon measurements. So the natural progression for photonics is very different to the conventional progression from single to two-qubit gates, each based on very different physics.  

In Fig.~\ref{fig:lattice} the depth in the $y$-direction is 6, and yet this lattice will, with high probability, have paths extending $O(10^4)$ qubits in the $x$ and $z$ directions. We see therefore that the percolation is remarkably efficient - so far we have been unable to find a 2d lattice that we can percolate with such small initial resources as 3-photon GHZ states, and yet we need go only a very small way into the third dimension to get large scale entanglement.

The $x$-direction is the one across which our logical qubits will be defined. Because the amount of 3-dimensionality required purely for the purposes of percolation is very small, if the goal is to use Method 1, as per the original ideas of producing percolated cluster states \cite{Kieling2007}, then this suggests a `squashing down' of the third dimension onto a 2d chip would not cause too many problems. That is, it will only require a fixed number (about 6) of crossings for photons to reach the partner they need to interfere with, and exchanging photonic qubits these sorts of distances is not an issue. (It could be argued that photonics should be looking beyond local, lattice-based codes because there are not the same spatial locality constraints as for matter qubits - for example there are interesting finite-rate codes \cite{terhal15} unavailable to `locality-contstrained' qubits). This architecture would consist of sources at the one end of the chip, interferometers in the middle and detectors at the far end. I will call this a 1+1 (one space, one time) dimensional architecture. However; if, as seems likely, we would rather follow Method 2, then a 1+1 dimensional architecture is more problematic. Fault tolerance is achieved in the surface code lattice by having a large amount of qubits in the $y$-direction. Doing the same kind of lattice squashing would require the non-locality of the photonic interactions to grow as the size of the computation grows.

It turns out that it is possible \cite{mercedesthesis} to engineer a 2-dimensional photonic chip  to allow for arbitrarily large 3-dimensional logical cluster states, in such a way that the number of photon crossings required is at most 1. It is this architecture that will be overviewed in the next section.   

Here are some of the theoretical questions which remain largely unexplored on this particular topic:
\begin{myenumerate} 
\item What are good protocols for exploiting serendipitous loss tolerance?
\item Ultimately should we pursue Method 1 or Method 2 or some other variant/combination? What are the best `path finding' algorithms for each case? What are minimal computational windows required in each case?
\item How local can the classical side-computation of measurement basis be? In Method 1 we believe it can be highly local, as for standard cluster state computation on a non-percolated lattice, but for Method 2 much less is known. 
\item Recently it was shown that almost all the well-studied quantum error correcting codes can be ``clusterized'' into a 3d lattice \cite{StacePrivateCommunication} very similar to the surface code approach. Should we be using one of these variants, particularly as we are more concerned about loss tolerance than stochastic noise anyway? Should we give up on spatially local codes altogether?
\item Photons need not be qubits, they can readily encode higher dimensional systems, which, at least abstractly, can be advantageous \cite{earlhigherdimensions,Lanyon:2009}. Can this be translated into a practical advantage? 
\end{myenumerate}

\section{Physical overview of the architecture}\label{sec:physicalarchitecture}

\begin{figure}
\centering
\includegraphics[width=\linewidth]{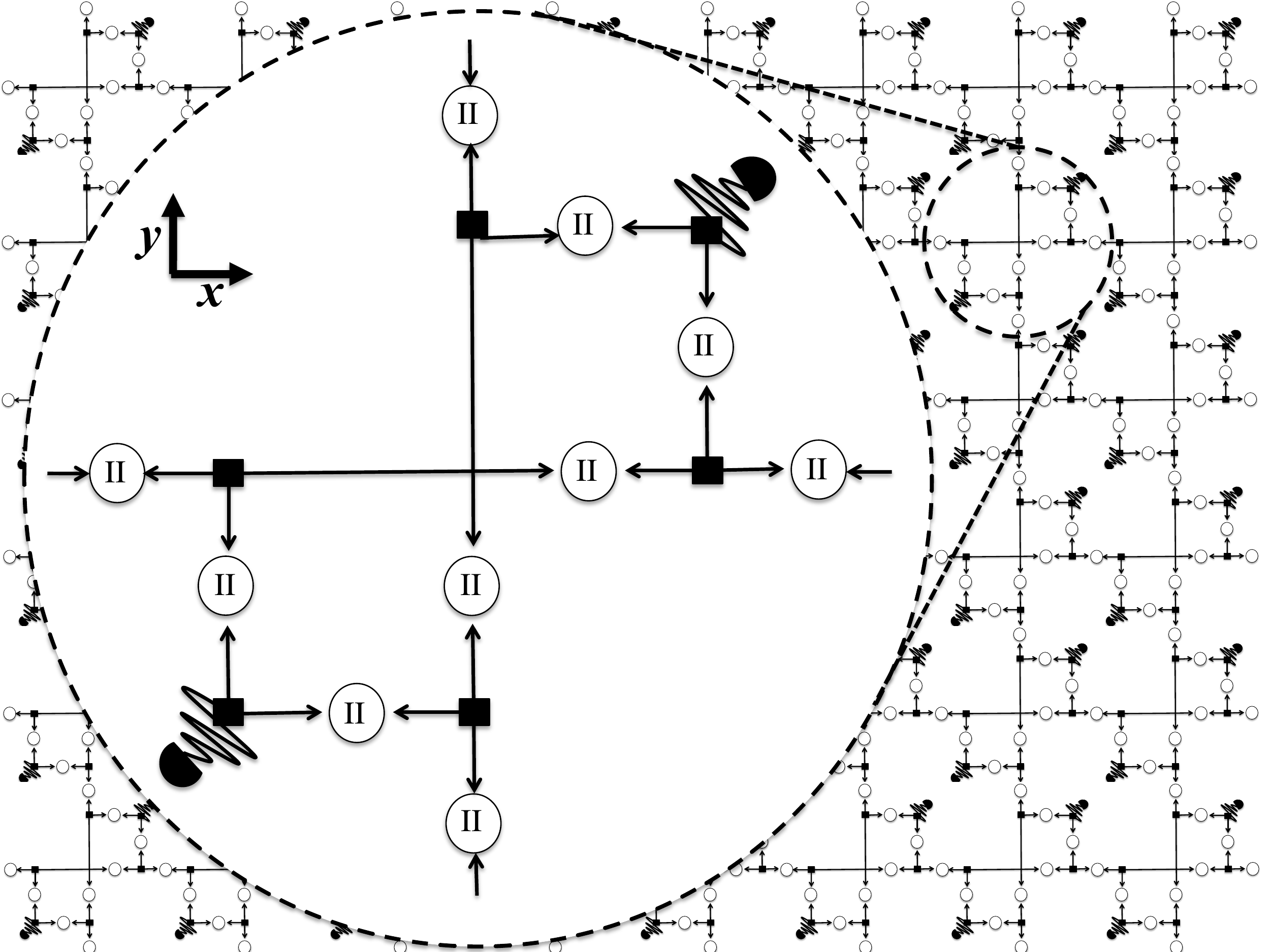}
\caption{Unit cell of the wafer which extends in the $x$ and $y$ directions of the logical lattice of Fig.~\ref{fig:lattice}. Circles labelled ``II'' are boosted Type-II fusion gates. Black squares are sources, each emitting a 3-photon GHZ state. Two of the 18 emitted photons need to be delayed before measurement (the time delay corresponding to the $z$-direction of Fig.~\ref{fig:lattice}), these are computational photons and are the only ones undergoing an active phase shifter once the photons are emitted from the source. Note that the only crossing here is that of the two photons at the centre of the unit cell.}
\label{fig:unitcell}
\end{figure}

The big-picture view of the 2+1 dimensional architecture (see \cite{Gimeno-Segovia2015,mercedesthesis} for many more details) is depicted in Fig.~\ref{fig:unitcell}. We imagine a large semiconductor wafer, probably silicon, divided into a square lattice. The two dimensions $x$ and $y$ of the wafer correspond to the $x$ and $y$ dimensions of the logical cluster state depicted in Fig.~\ref{fig:lattice}. 

Each unit cell of the wafer contains 6 sources capable of producing 3-photon GHZ states (equivalent to a 3-photon linear cluster state) with high efficiency. Of these eighteen photons, two will be computational qubits in the final logical lattice. One of these will be ahead of the other, i.e. in Fig.~\ref{fig:lattice} they will lie along the line of qubits that extend into the $z$-direction emanating from each $x,y$ location (for aficionados of surface code cluster state - one qubit is primal and one is dual). A delay (around 10-15 clock cycles due to the computational window mentioned above) is necessary before we know which basis these photons will be measured in. This is depicted as a winding waveguide delay (see Sec.~\ref{sec:delays}) before the detector. Of the eighteen emitted photons it is only these two which pass through an active element (a phase shifter to choose the measurement basis). We call this architecture `ballistic' for hopefully obvious reasons. 

Four of the sixteen non-computational photons are used to connect the unit cell to adjacent cells, by entering a `Boosted Type-II' fusion interferometer at the cell edge. Type-II fusion \cite{Browne2005} is an operation that with probability 1/2 joins up (fuses) cluster state, the boosted form \cite{Gimeno-Segovia2015} exploits the ideas of \cite{Grice2011,Ewert2014} to increase the success probability to 3/4. The boosting process itself consumes either four single photons or a Bell pair, so within each such `II' box indicated in the figure more sources are implicitly necessarily integrated on the chip. The remaining twelve non-computational photons all undergo fusion gates with each other within the cell.                                                                                                      

Note that each photon's world line in this architecture is really short. Even including the non-deterministic circuits to convert single photons into the 3-GHZ states, the `optical depth' of any given photon (the number of elements it passes from birth to measurement) is on the order of 10, and the number of other photons it ever needs to interfere with is at most 3. Contrast this with boson sampling \cite{Aaronson2011b} which requires optical depths of several hundred elements and interference with many tens of other photons. Perhaps the easiest way to do boson sampling (though why would you?) is to build a photonic quantum computer!

In terms of the initial 3-photon GHZ generation, coupling in photons from off-chip sources via fiber is possible, but this is clunky and ultimately hopefully avoidable. The sources will need to operate on some clock cycle that is at least in the megahertz, but hopefully gigahertz range. 
I should say that although I lean towards being `as monolithic as possible', excellent photonics engineers have expressed the sentiment to me that they envision an architecture like this as consisting of a large number of separate modular chips talking via low-loss interconnects.  
Either way what is important is that this architecture avoids multiple stages of massive, technologically demanding active switching and routing networks throughout the computation to construct complex resource states - without doing so the resource scalings would be prohibitive \cite{Ying15, Lovett2010}.

The last thing to point out about the architectural schematic envisaged here is that a 2+1 configuration will likely require pump laser distributed around the chip, and will also have some regions of the wafer devoted to classical electronics for the computations that determine the measurement bases. These may also be implemented in different layers \cite{Koonath:07}. As mentioned already, ideally such computation will use simple logic and be local. But to some extent the simplicity of this whole architecture arises because we have shunted into a classical computation much of what we used to need to deal with actively in a tricky quantum-coherent way. So perhaps I'm just being greedy to want the classical computation to also be simple and local, and yet I am reasonably sure that such is feasible. 
 
 To summarize, if we had highly pure sources of 3-photon entanglement, compatible with CMOS fabrication, then our photonic architecture becomes incredibly simple: our overhead would be 16 physical photons per final qubit in the cluster state and only two photons would undergo any kind of potentially noisy active element. The rest of the computer is just an interferometer built from components that are already able to be fabricated to tolerances sufficient for us to generate large amounts of entanglement.   
 

Here are some of the theoretical questions which remain largely unexplored on this topic:
\begin{myenumerate}
\item Our current proposal is to use interferometers that are loss tolerant, in that we only herald off registered clicks (another improvement over the original percolation proposal \cite{Kieling2007}). However, we know we can give up such loss intrinsic tolerance for higher success probabilities (e.g Type-I fusion \cite{Browne2005}), which leads to more connected lattices etc. Is this beneficial?
\item We have no proof that 3-GHZ initial states are necessary. Perhaps ballistic protocols with Bell pairs, or even single photons are possible, particularly if we take into account options, as per the previous question, which do not destroy two photons at a time? As far as I know only one, not-particularly-smart, theorist has devoted about two days to this question. 
\item Is it possible to boost fusion to higher probabilities (with ancillas that are no more entangled than 3-GHZ states)? We have no methods (even numerical) for verifying the optimality of \emph{any} of the current interferometric gates we use! Forget playing Go \cite{deepmind}, can't we get some artificial intelligence (other than Mercedes Gimeno-Segovia) onto this problem?
\end{myenumerate}

\section{Detectors}\label{sec:detectors}

Ten years ago poor detectors would have sat alongside sources as the primary obstacle to photonic quantum computing. With the advent of number-resolving, fully integrable superconducting detectors \cite{marsili2013detecting,akhlaghi2015waveguide},  the problem is well on the way to being fully solved. 

Efficiency percentage in the high 90's, jitter in the low 10's of picoseconds, reset times on the order of 10's of nanoseconds and and vanishing rates of dark counts are achievable. It should be emphasized, however, that even very small improvements in detectors can lead to large savings downstream in terms of the overall scale of the computer. This is particularly true if our sources are of a multiplexed variety. 

There does not seem to be much interesting a theorist can investigate about detectors per se, so we just encourage our experimental colleagues to make them better. Oh - and getting them to work at room temperature would help make lab tours for theorists more interesting.

\section{Switches/phase shifters}\label{sec:switches}

A phase shifter in a Mach-Zender configuration can act as a switch. Phase shifters can be built on a thermo-electric effect, but these are slow and in general we expect they will need to be built on some variation of an electro-optical effect, or, if speed of operation is not crucial, on an electro-mechanical effect. Nanoscale electro-mechanical devices are slow (though megahertz achievable) but do have the advantage that they are extremely low loss when in the `open' position, unlike those based on electro-optical effects. 

 There are a very wide variety of such devices being investigated (within classical photonics engineering), and I am not well-versed enough to summarize them all. There seem to be no good physical reasons that these types of switches cannot have large performance increases in the near future. There will, however,  always be tradeoffs: at present, in silicon, they can be very fast (e.g. gigahertz \cite{826874,liu2004high}) but then are lossy; they can be slow (e.g. megahertz \cite{Seok:15}) and very low loss, and a wide variety of performance characteristics in-between. 

As switches and phase shifters are the primary source of stochastic noise it is worth understanding some of the limiting factors in switch performance. Static limitations such as fabrication tolerances, accuracy of the voltage source, polarization/spectral-width/other wavepacket issues and so on lead to imperfections, as discussed in Section \ref{sec:photonsasqubits}, which can be well characterized and dealt with much more easily than stochastic noise. The dominant source of stochastic noise (at cryogenic temperatures drowning out those from induced refractive index changes due to thermodynamic fluctuations/vibrations etc) is expected to be electrical noise (thermal, shot noise, $1/f$ etc.) on the voltage source.

It should be pointed out that if we do have a source of high quality GHZ states then the photons will (at most) have to pass through one switch/phase shifter before measurement. In particular, computational photons will need to be measured in either the Pauli-$X$ or $Z$ bases if we take a surface code + magic states approach. The former amounts to interfering the two modes that comprise the qubit on a 50:50 beamsplitter, the latter to just measuring the two modes directly with no interference. One option is to send the two modes through a Mach-Zender and adjust the relative phase in the paths to 0 or $\pi/2$. 

An alternative is to use a switch to direct the two photonic qubit modes toward some static measurement setups - i.e. either direct them straight toward some detectors, or direct them toward a static 50:50 beamsplitter followed by detectors. The latter method has a possible advantage in that we can characterize and then tune the static imperfections of the effective $X,Z$ measurements very well. This ``set-and-forget'' possibility has already been demonstrated to work with incredible accuracy \cite{Shen:11, Shen:11b}. Also, since $X$- measurements are more sensitive, perhaps we make them the default and switch only to a $Z$-measurement? More interestingly, from a theory perspective, things can be arranged so that if the switch does not operate perfectly, in as much as the photon goes the wrong way, all that happens is that we obtain a measurement in the undesired basis. But it is presumably better (and algorithmically correctable) to do a good clean measurement in the wrong basis than measure in the right basis but get the wrong outcome because of stochastic error. So if, as seems likely, demanding less perfect performance from our switches can reduce stochastic error then this is worth investigating.

The upshot is that the stochastic noise in the switches is probably at worst going to affect the relative phase between the two paths being directed towards the 50:50 beamsplitter for the $X$-measurement. The fact that this error is primarily of Pauli-$Z$ type is also of relevance once we consider quantum error correction and fault tolerance more abstractly. There may well be fancy engineering tricks to minimize this. For example, perhaps we can drive both switches from the same voltage in such a way that first order fluctuations cause opposite sign phase shifts (thats just a vague theorists idea!) or use a NEMS that has this measurement as its open-position static default. But it would be nice to get an estimate of how low we can make this noise with current technology before worrying about future possibilities.  

Despite persistent nagging on my part, very little experimental effort has been made to try and obtain an accurate direct measurement of how low the stochastic noise can be made in photonics. The numbers are small, so it is difficult, but presumably by suitable cascading of devices (and perhaps incorporating randomized benchmarking techniques \cite{knillrandomizedbenchmarking}) the effect can be amplified. We can, however, obtain an indirect estimate. It is currently possible to build a Mach-Zender type switch \cite{Suzuki:15} with an extinction ratio of about -50dB. If the imperfectness of the extinction were completely due to a random phase shift $\Delta\varphi$ in the arms of a perfectly balanced interferometer (it most certainly isn't, its mainly due to static imperfection!) then we would say $\sin^2\Delta\varphi\approx(\Delta\varphi)^2\approx 10^{-5}$. Such a phase fluctuation in the $X$ measurement is equivalent to having a Pauli-$Z$ error rate of $10^{-5}$. But it would be surprising if we cannot get down to around $10^{-8}$, which is roughly what shot noise at the 1 volt, 1 milliamp range would imply. Of course I am then told that sub-shot-noise operation is possible, but I have no idea at present how low that might push these numbers.  \footnote{\emph{Hot off the press:} While finalizing this article I received some preliminary data \cite{newbristolexperiment} from an experiment at the Centre for Quantum Photonics in Bristol, where this extinction ratio has been improved to around -65dB.}

Here are some of the theoretical questions which remain largely unexplored on this topic:
\begin{myenumerate}
\item Are there advantages to using different speed switches for different roles within the full architecture?
\item Is there really an advantage of `set-and-forget' over Mach-Zender type choices of measurement basis?
\item Should we aim for lower performance switches to reduce stochastic error and try to algorithmically correct incorrect measurements being performed?
\end{myenumerate}

\section{Delays}\label{sec:delays}

Photonic quantum memory is a delicate technology (see e.g. \cite{WalmsleyQuantumMemory}), requiring controlled interaction of light with matter. It is crucial to the viability of photonic quantum computing that this ballistic architecture does not require such memory, because at present their performance is very far from what we require, and they are not integrable. Also, by involving matter we would need to be concerned about the addition of stochastic error within the memory. Of course if high-performance memory does become available (and integrable) it would likely be extremely useful.  

All we need for this architecture are fixed time delays. Delaying a photon a fixed number of clock cycles involves only a long, passive, waveguide. On-chip delays that basically wind a waveguide into a compact spiral have already been constructed \cite{lee2012ultra,Li:12} with lengths up to  27 meters! If we have 100 megahertz switches this is already longer than we require. Loss occurs as the photon travels around bends, and these can be greatly ameliorated with tricks that expand and constrict the waveguides in judicious ways. It is not unreasonable to expect delays with losses of less than 0.01dB per meter \cite{lee2012ultra}.

\section{Sources, multiplexing and all that}\label{sec:sources}

Methods for producing single photons have been under active investigation in a wide variety of systems. Given 6 single photons it is possible to produce a 3-photon GHZ state, but only with probability 1/32 \cite{Varnava2008}. In a perfect world we would be able to produce 3-photon GHZ states directly. This certainly can be done, for example the photon machine gun \cite{Lindner2009} has recently been built using a self-assembled quantum dot in the Gershoni group \cite{gershoni}. Many other options for building the photon machine gun exist - NV centers in diamond, trapped atoms and ions are all technologies for which the current method of producing single photons is more or less amenable to producing 3-photon GHZ states instead (though admittedly it is more complicated). The extent to which any such method is compatible with being implemented directly on chip varies greatly, as do achievable repetition rates, as does the indistinguishability of the photons produced from different sources. The photons could be produced off-chip and coupled in by fiber; present state of the art for such coupling is an efficiency of -0.36dB \cite{Notaros:16}. Most concerning, however, is that these methods are vulnerable to stochastic noise being transferred into the photons from the matter responsible for producing them.   

\begin{figure}
\centering
\includegraphics[width=0.8\linewidth]{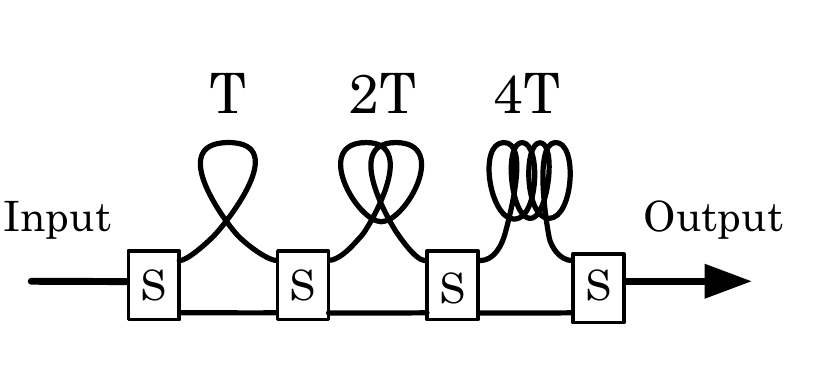}
\caption{Cascading delays of length 1,2,4,8,... allows $S+1$ switches to delay a photon by any time between 0 and $2^S-1$. The figure makes it look as if the delays are in fiber, in reality they would be waveguide `spirals' as in Section~\ref{sec:delays}.}
\label{fig:switches}
\end{figure}

A technology we have considerable experience with that can produce highly pure, single photons is spontaneous parametric downconversion (SPDC), where a single photon is heralded via the detection of its partner. Separate SPDC sources can produce highly indistinguishable photons \cite{LaingHiFi}. The states from a single SPDC source are highly pure, it should be said, except for the time-bin degree of freedom - the emission time is random. Roughly speaking such a source can produce single photons at 10's of gigaherz repetition rates, with a (controllable) probability of any particular time bin containing a photon ranging from about 0.1\% to 20\% (we assume efficient detectors herald any multi-photon emissions, which are then ignored). Particularly relevant for our purposes is that the spontaneous four-wave mixing variant (SFWM) of such nonlinear optical photon production is fully compatible, in fact already manufacturable, within an integrated architecture - it does not even require any new type of component to be introduced to the fabrication process. This is because we can use the small nonlinearity of the semiconductor itself and obtain heralded single photons by pulsed pumping of long waveguides, or of waveguide cavities \cite{spdconchip,Gentry:15,spring16}.  

The key issue for such sources is to produce photons that are highly indistinguishable. Typically the photons pairs produced (which may be quite different frequencies) will be entangled, in which case we need to measure the heralding photon to collapse its partner onto a pure state. Doing this measurement requires strongly filtering the heralding photon to ensure high purity of the partner, and although extremely good filters exist such filtering decreases the probability per pulse of the photons. Alternatively sources which pump cavities directly produce the photon pairs in a product state using high-quality waveguide cavities. Typically indistinguishability is measured in terms of visibility of a HOM test \cite{HongOuMandel}. Although extremely high visibilities of the photon pairs produced in a single nonlinear process are possible (showing how strong the correlation/entanglement is), we cannot expect nearly as high visibilities from heralded photons produced from two different such processes. My suspicion is that non-unit visibility arising from two photons being in slightly different pure states is much less detrimental than that caused by the photons being slightly mixed, but a detailed investigation needs to be done. Regardless of how pure we produce these heralded photons, they are still produced by a spontaneous process and so emerge in random time bins.

A nice idea \cite{Migdall2002} to remove the time-bin randomness is that of \emph{multiplexing}. This can take either a spatial form - place many sources in parallel and use a switching network to switch out a photon from a source you know has fired, or a temporal form - use adjustable delays on a single stream of photons to shuffle them backwards into desired time bins. Both methods require the photons to traverse switches, which are both lossy and our only significant source of stochastic noise and so minimizing switching is the key concern. 

Most of what I will say here is generic, but I will use temporal multiplexing for definiteness. We have a stream of photons where each time bin contains a photon with probability $p$. Using $S+1$ switches a photon can be delayed by any amount of time between 0 and $2^{S}-1$, as depicted in Fig.~\ref{fig:switches}. What I will refer to as \emph{standard multiplexing} divides the stream of random photons into blocks of length $2^S$. If a block contains a photon it shuffles that photon to the back of (i.e. the last time bin in) the block. If there is more than one photon in the block then the extra photons are ignored/discarded. The final stream of photons can then be viewed as having a new time bin period of $2^S$, with a new probability of a photon in each bin of \begin{equation}
p'=1-(1-p)^{2^S}. 	
    \end{equation}
    If the switches were lossless this would let us readily turn our non-deterministic source into an effectively deterministic one.

This would not, however, be the end of the need for multiplexing. We need to turn these deterministic single photons into 3-GHZ states or Bell pairs. The methods we use for doing so are probabilistic   \cite{Zhang2008, Varnava2008, mercedesthesis}. As such we need to multiplex again after the probabilistic production of the desired states.

\subsection{Asynchronous computation  and relative-time multiplexing}

It turns out that large improvements are possible to standard multiplexing. The big-picture realization is that there is no reason the photonic quantum computer need be run perfectly synchronously - an asynchronous operation will do. The whole reason we are trying to reduce the indeterminism in the source is that any particular photon is going to need to arrive at some optical element at the same time as some other photon with which it will interfere. But it certainly need not arrive at the same time as all other photons in the quantum computer. This leads to the concept of \emph{relative time multiplexing} (RMUX).  
    
\begin{figure}
\centering
\includegraphics[width=0.9\linewidth,bb=5 5 360 270,clip=true]{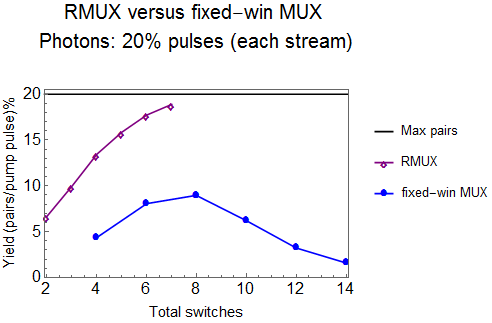}
\caption{Comparison between relative time (purple) and standard (blue) multiplexing for two streams of photons, each with 20\% probability of a photon per pulse, where photons from one stream are delayed in order to `partner up' with photons from another. The yield is the final percentage of pulses that end up with two photons impinging on a beamsplitter. The yield in standard multiplexing eventually drops because only one of the many photons produced in a large standard multiplexer actually gets used.}
\label{fig:RMUX}
\end{figure}

Full RMUX is a bit complicated, so first consider a simpler ``sliding window'' method that captures some of the idea. Imagine we want to impinge two photons on a beamsplitter. We have two streams of photons in (known but) random time bins and each photon can be delayed up to $2^S-1$ bins. We consider the first incoming photon from either stream, and ask whether there is a photon from the other stream that is no more than $2^S-1$ bins behind the first. If so then we delay the first photon appropriately to meet it, and repeat. If not we discard the first photon and look to the second incoming photon from either stream. The expected number of `matched photon pairs' can increase significantly because we are not forcing the photons into a particular clock cycle, rather we are only adjusting indeterminism in their relative time separation. 

In practice we would use the generalization of the sliding window idea to either four streams of photons to produce Bell pairs \cite{Zhang2008},  or to six streams \cite{Varnava2008} to directly produce the 3-GHZs we so desire. We also will likely only put delays on one of the photon streams to reduce the number of switches. But the sliding window technique still discards many potentially useful photons, because only one photon per stream within the current window is used. At low values of $p$ this is not significant, but at higher values we are failing to use many photons. Therefore it makes sense to look at more sophisticated \emph{matching} algorithms (well studied in graph theory), and these do significantly better again. The simplest comparison for the simple case of two photon streams is shown in Fig.~\ref{fig:RMUX}. 
There is a subtlety however; when the density of photons is high there is the potential for `collision' of photons within the delay network, and these must be accounted for. 

Where RMUX really comes into its own is when we have the streams of 3-GHZs that we are now fusing to produce our final logical lattice, as overviewed above (details of the fusing configurations are in \cite{Gimeno-Segovia2015}). Non-computational photons within each unit cell (i.e. the majority), which effectively live only to produce all the bonds of the final logical lattice, need not synchronize with any photon other than the one with which it undergoes fusion (there also being photons within the boosted fusion gate to consider). This saves considerably on the switching  and delays required. The whole computer operates asynchronously - but it is one of the wonders of quantum entanglement that the ordering of events on different members of an entangled state is irrelevant to the quantum correlations (despite quantum nonlocality as it were), and we can utilize this to great advantage. Moreover, different photons (typically the computational qubits) can be judiciously chosen to end up having to go through many fewer switches than others. The whole procedure is quite complicated and still undergoing considerable optimization.

Here are some of the theoretical questions which remain largely unexplored on this topic:
\begin{myenumerate}
\item Are there better delay networks, particularly ones designed to avoid collisions?
\item What are the best combinations of matching and multiplexing procedures, given we do not need synchronous GHZ states? 
\item How much advantage can we gain from asynchronous operation, taking into account the fact we cannot have arbitrarily long delays before measuring computational photons?
\item Not all photons have to go through the same number of switches, so which apportionment of switching asymmetry is best?
\end{myenumerate}

\subsection{Dump the Pump?}

One of the reasons our switch technology needs to be so good is that our photonic qubits are interacting directly with the switch, and qubits are delicate creatures. Interestingly this is not, in fact, necessary. A different way to multiplex heralded photons is to turn off the pump instead. 

\begin{figure}
\centering
\includegraphics[width=0.9\linewidth]{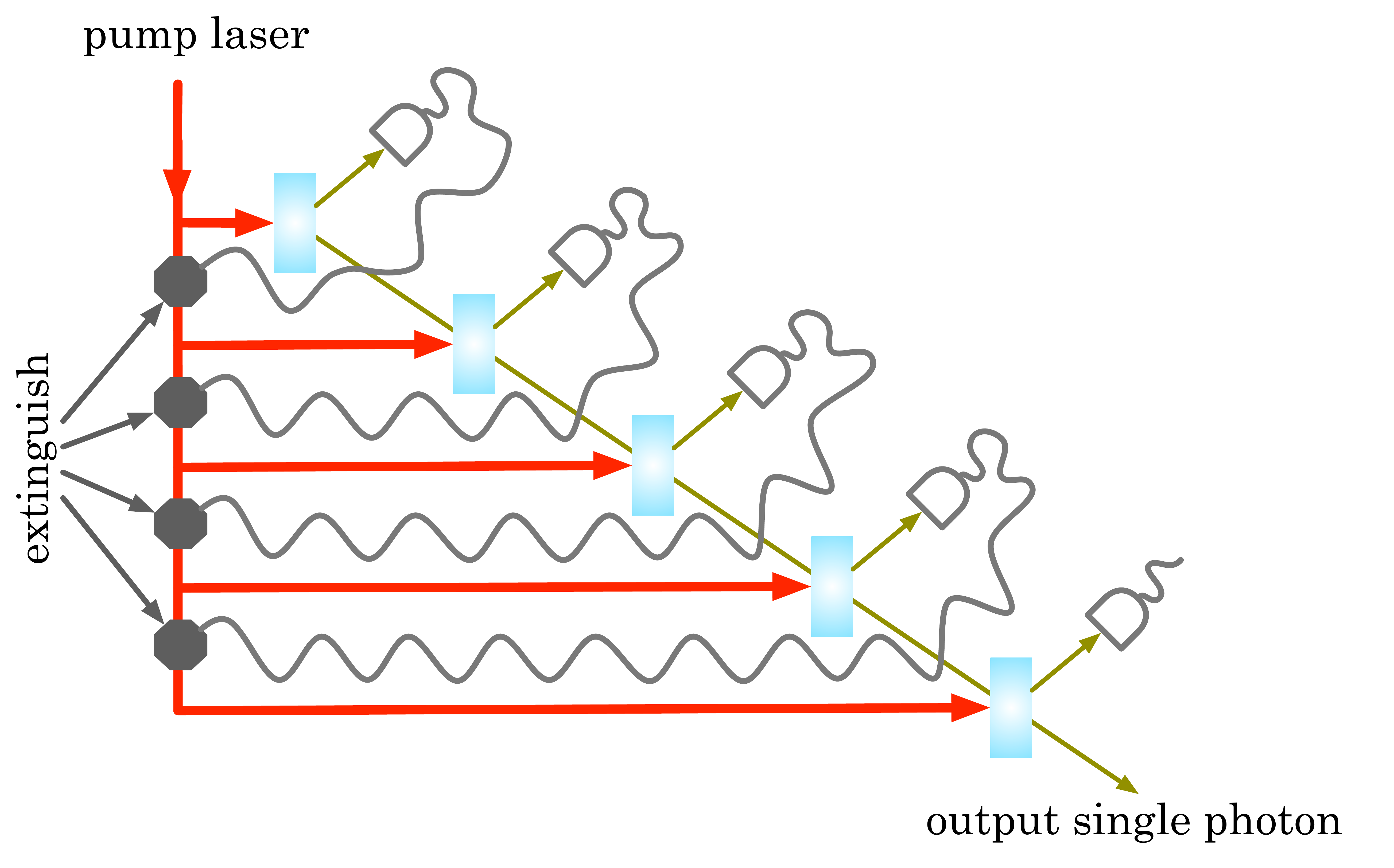}
\caption{Dump-the-Pump: A way to avoid switching of our photonic qubits.}
\label{fig:DTP}
\end{figure}

The idea is easiest explained in a bulk-optics SPDC schematic although this is not, of course, how we would implement it on-chip. Recall the famous two-crystal experiment of Zou, Wang and Mandel \cite{Zou1991} where two nonlinear crystals were aligned and pumped in such a way that it was impossible to tell if a detected photon had originated in the first crystal or the second. The idea of dump-the-pump is to cascade such a process, but rather than do some fancy interference experiment with the idler photons as Zou et al did, we simply use them to herald when a photon has been produced (see Fig.~\ref{fig:DTP}). If a photon is detected then we turn off the pump to all the remaining crystals. We view the Zou et al experiment as proving the important point that this process can produce indistinguishable photons.

The nice thing about dump-the-pump is that the potentially noisy active element is operating on the pump. Not only is the frequency of the pump very different to that of the photonic qubits, it is only necessary to extinguish the pump not to coherently switch it per se. This opens up quite different switching devices to those currently considered for multiplexing. The disadvantage is that once a photon is produced it needs to pass through all the crystals, a potentially lossy process. The procedure is also `linear depth', you don't get the compactness of a logarithmic depth switching network as in Fig.~\ref{fig:switches}.          

At a high level one can think of dump-the pump as effectively taking a long nonlinear crystal and slicing it, so that we remove the uncertainty as to where exactly the photon was produced. From one strongly pumped SPDC source we could at most hope for a probability of single photon emission around $1/5$. I suspect dump-the-pump could be most useful to raise this probability to around $2/3-3/4$ (requiring 5 or 6 cascaded downconversions) at which point multiplexing the single photons as discussed above becomes highly efficient and possibly the better way to progress.   

Here are some of the theoretical questions which remain largely unexplored on this topic:
\begin{myenumerate}
\item SPDC can be used to produce heralded Bell pairs and GHZ states not just single photons; is there an advantage to using dump-the-pump (or regular multiplexing for that matter) on those setups?
\item How efficiently and how fast can we extinguish a pump? How do we optimize the likely combination of dump-the-pump and RMUX, given estimates of switch performance parameters attainable in the near future?  
\item We tend to focus on single photon encodings of our quantum information. But good number-resolving detectors means we can have heralded sources of higher number Fock states, possibly with higher efficiency than single photons (imagine dump-the-pump without switching off the pump). Two such modes containing $N$ photons in total are equivalent to a spin-$N+1$ particle, which can be easily evolved under the corresponding representation of $SU(2)$. Little is known about the computational power of such systems - is `photonic universality' easily achieved in such encodings?
\end{myenumerate}

\section{Loss tolerance}\label{sec:losstolerance}

The most important error mechanism to affect photonic quantum computing is photon loss, which can arise at any point during the photon's journey. Unlike stochastic error, however, loss error is known to have occurred when a particular detector fails to register a photon. Crucially, by making use of variations of the Type-II fusion gates only \cite{Browne2005}, it is possible to design the architecture to only make use of events that trigger off photon clicks. Depending on the details these are not  even necessarily from number-resolving detectors.  
 
\begin{figure}
\centering
\includegraphics[width=0.9\linewidth, bb= 60 100 650 490,clip=true]{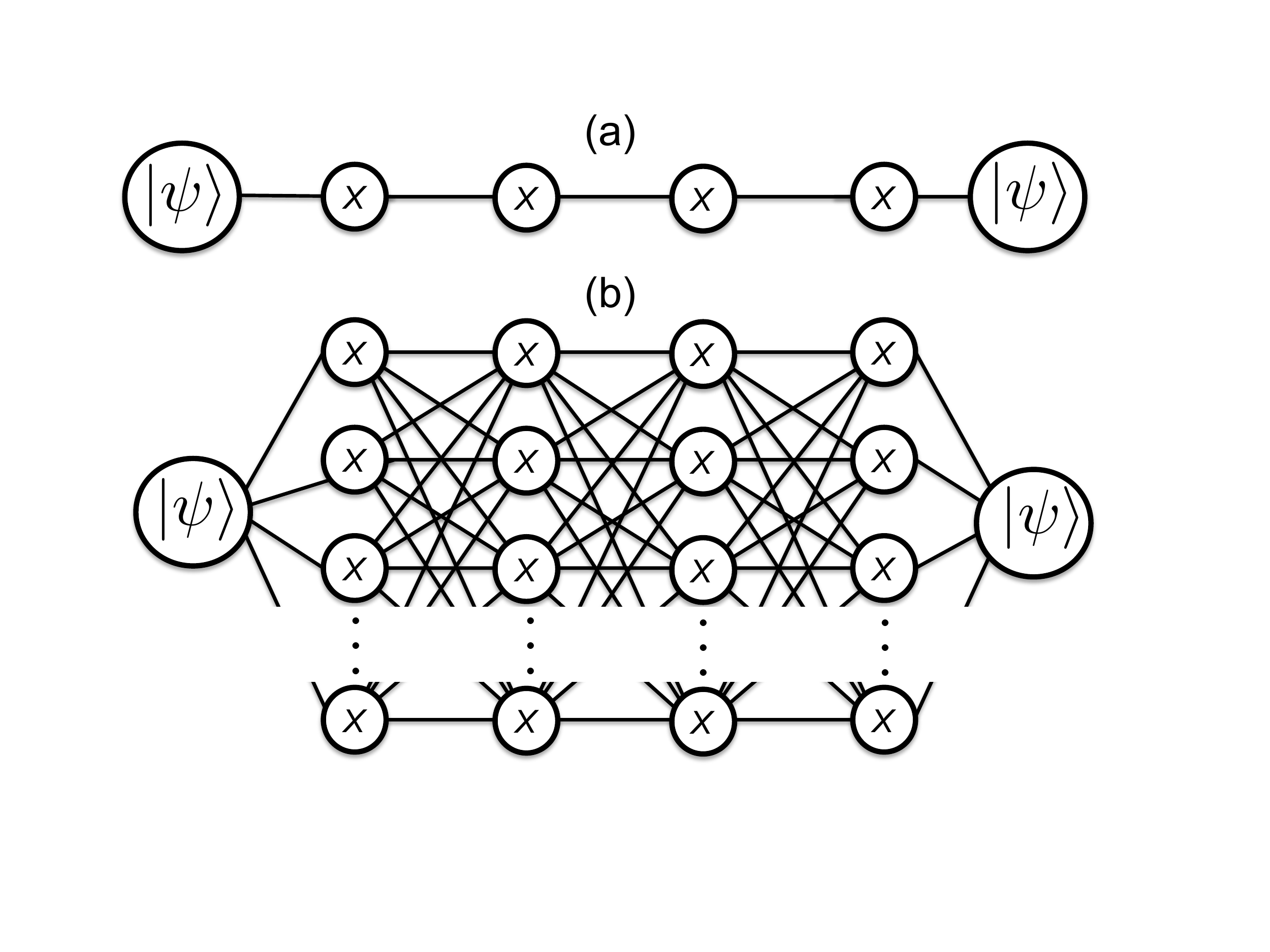}
\caption{Teleportation along a linear cluster state (a) and its loss tolerant ``crazy graph'' encoding (b) where each qubit of the linear cluster is replaced by a column of $L$ qubits.}
\label{fig:crazygraph}
\end{figure}

Two mechanisms for dealing with losses have been mentioned so far: punching out loss by measurement and logical renormalization to super-stabilizers. Both are fortunate byproducts of some unrelated architectural flexibility, but in reality we should aim to build in loss tolerance explicitly. Understanding how to do this within a ballistic architecture such as the one discussed here is ongoing work. Although I cannot yet provide a mechanism for very high loss tolerance in this particular architecture, in this section I want to overview why I am optimistic about loss in general being an extremely benign error mechanism. In fact I will argue that in the absence of stochastic noise anything less than $100\%$ loss is in principle tolerable!      

It was shown in \cite{Varnava2006,Varnava2007} that there exist cluster states (`tree codes') for which the overall loss probability per photon can be as high as $50\%$. There are several disadvantages to this scheme. It requires quantum memory (itself lossy) in order to build the trees using probabilistic gates. Also the trees themselves involve very large numbers of qubits when the loss rate is high. To some extent these disadvantages have been reduced by magic state injection methods, because the tree codes were designed to work in the presence of both high loss and arbitrary non-clifford measurements. However, with magic state injection techniques \cite{Bravyi2005} the measurements need only be Pauli-$X$ and $Z$ basis measurements, and the whole tree structure is designed to allow inference of these outcomes counterfactually when a photon is lost. Thus the restriction to $X,Z$ measurement can reduce the size of the trees required significantly. It forms the basis of a recent quantum repeater proposal \cite{LoQuantumRepeater}.

If we want to protect against loss there is a different method, particularly suited to magic state type protocols. For the basic idea, consider teleporting an arbitrary state $|\psi\rangle$ along a linear cluster state ``wire'' as in Fig.~\ref{fig:crazygraph}(a) via $X$ measurements. Losing even one qubit causes the propagation to fail. However if, as in Fig~\ref{fig:crazygraph}(b), we replace every qubit of the original wire by $L$ qubits, completely connected to its neighbors as shown, then as long as at least one qubit per column survives, the teleportation will succeed. 

\begin{figure}
\centering
\includegraphics[width=0.8\linewidth, bb= 60 180 550 460,clip=true]{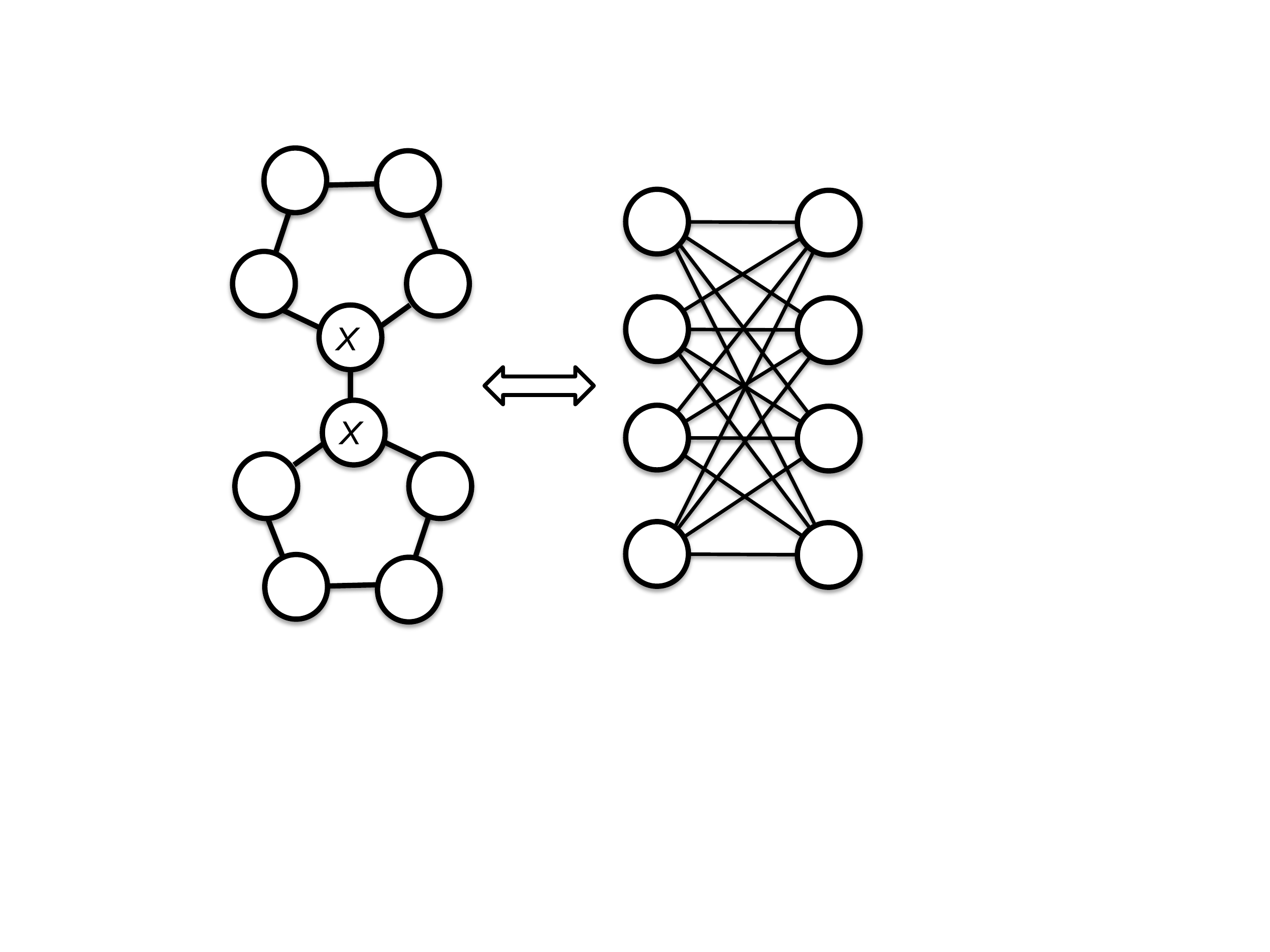}
\caption{Highly connected graphs do not necessarily involve as many CZ gates as it may at first seem. In the left hand graph each qubit has bounded degree. Measuring the two qubits indicated $X$ results in the right hand piece of crazy graph. This generalizes - increasing the number of qubits in the rings on the left increases the number of qubits in each column on the right.}
\label{fig:crazygraphequivalence}
\end{figure}

For any loss probability per qubit $\epsilon<1$ and wire of length $N$ we can choose `polynomially large' $L$  such that $\left(1-\epsilon^L \right) ^N \approx 1$. So in principle we can tolerate $100\%$ loss for this process! Of course as $L$ increases the number of bonds in the graph (that we call ``crazy graph'') also increases (polynomially). Therefore unless we are able to perform extremely low stochastic error gates to build this cluster state (as we do anticipate in photonics) we will run into problems. To some extent this overstates the problem however. Firstly, any $Y$ or $Z$ Pauli error on a qubit will flip the value of the $X$-measurement outcome on that qubit. However, in the absence of noise all the qubits in the column  would register the same value outcome for the $X$-measurement. Thus a simple majority vote amongst those qubits in a column which are not lost will reveal what the correct measurement outcome should be. Secondly, building a highly connected graph does not automatically involve using lots of quantum gates, tricks to do with graph operations like local complementation can be employed. An example of a building block for crazy graph built by measurement on a low vertex degree graph is shown in Fig.~\ref{fig:crazygraphequivalence}

\begin{figure}
\centering
\includegraphics[width=0.8\linewidth, bb= 50 100 630 370,clip=true]{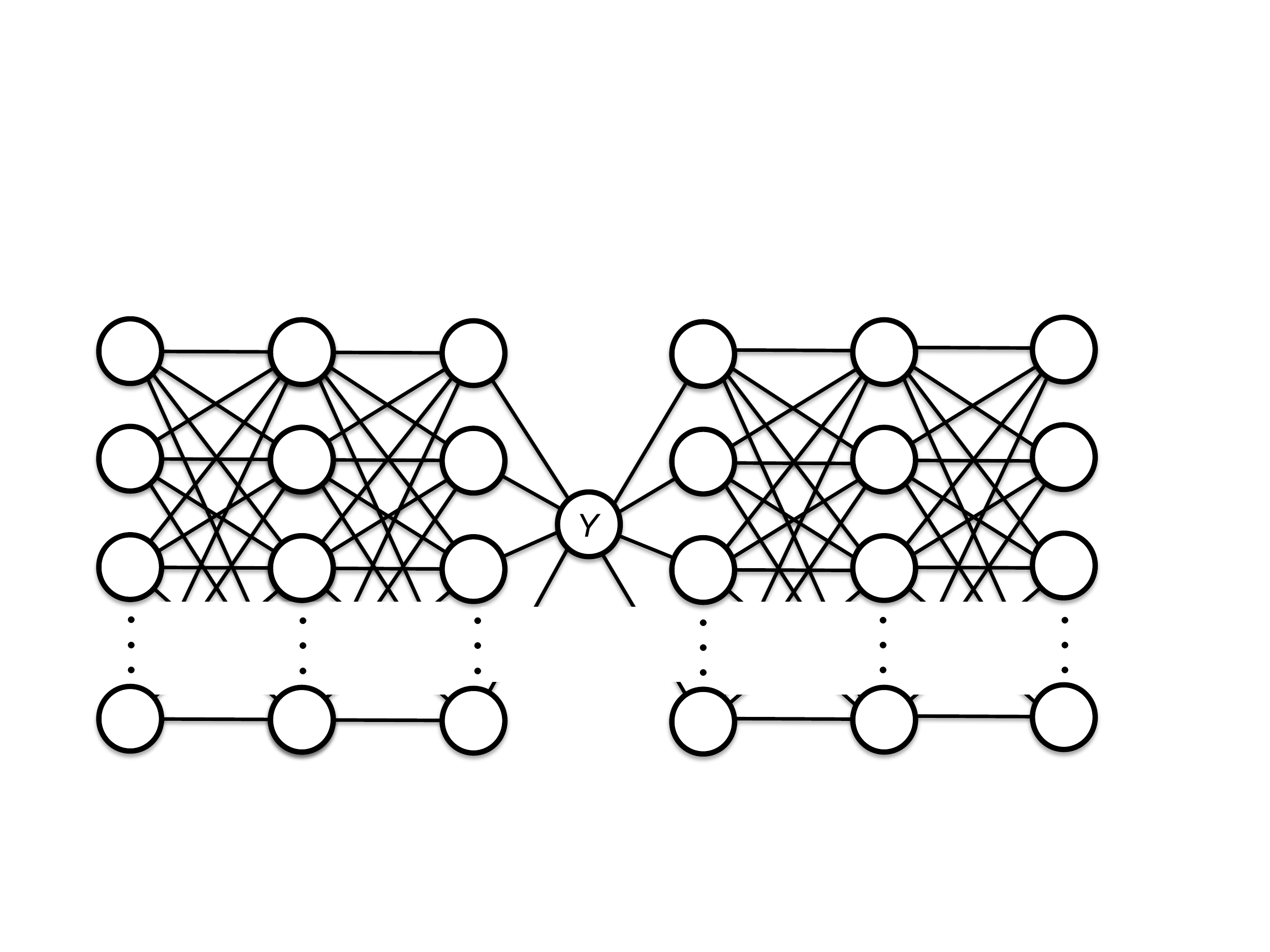}
\caption{A piece of cluster state that will implement the Clifford $S$ gate. The central Pauli-$Y$ measurement would be attempted before attachment to the computational cluster state, and the piece only retained if the qubit was not lost.}
\label{fig:crazygraphSstate}
\end{figure}

The loss tolerance for the wire can be extended for any clifford gate circuit we want to perform via cluster state methods. We imagine building up a crazy graph cluster state `on the fly' by attaching in chunks of pre-built cluster state just ahead of where the measurement based computation currently has reached. Hadamard gates are performed by simply changing whether the number of qubits in the underlying linear cluster is odd or even. An $S$-gate can be implemented attaching on a graph state pre-prepared as illustrated in Fig.~\ref{fig:crazygraphSstate}(b). Note that the central qubit measured in the $Y$-basis is not loss protected, however this piece of graph state will have been prepared in advance, and so cases where that qubit was missing will have been post-selected out. (Or, more naturally, we will have just buid the graph that results when the $Y$-measurement is performed).   

Similar pieces of cluster state can be built to do CZ gates between qubits and so on.

Here are some of the theoretical questions which remain largely unexplored on this topic:
\begin{myenumerate}
\item Is there a way to integrate highly loss tolerant code ideas into a ballistic architecture, without using quantum memory or extremely long delays? 
\item Crazy graph is tolerant to stochastic Pauli noise once it has been built, but are there mechanism for creating the graph that do not amplify such stochastic noise too drastically?  
\item Can crazy graph improve quantum repeaters, along the lines exploited for tree codes?
\end{myenumerate}

\section{Conclusions}

Since the seminal paper by Knill, Laflamme and Milburn \cite{Knill2001}, huge progress has been made towards an all-photonic quantum computer. On the experimental side we now have integrated photonics, including the detectors of our dreams ten years ago. On the theory side, gone is the circuit model and  gates that only worked with approximately unit efficiency even when using arbitrarily large numbers of photons, gone are the giant ancillary resource states requiring huge quantum memory to prepare offline, gone is the need for highly nonlocal operations... In short, gone are the eye-watering overheads which would forever have destined linear optical quantum computing to the fate of a mere intellectual curiosity, irrespective of the progress on the experimental front. Despite all this progress, it is clear that there are still a large number of potentially highly significant simplifications to the architecture for theorists to tackle.   

I have focussed here on a purely photonic implementation. However the architecture has  obvious potential for hybrid light/matter approaches, as well as networks of PICs with lossy interconnects \cite{nickerson1,nickerson2}, and these should be investigated further. 

Finally, given that many of the performance characteristics required can also already be met in optical fiber technology, and switches there can be significantly lower loss \cite{KumarSwitch}, there is some temptation to imagine an all-fiber photonic quantum computer being built in a disused aircraft hangar, with huge spools of fiber delays hanging off walls and ceilings - a Quantum ENIAC as it were. As with the original ENIAC, I suspect this possibility will only be explored seriously by the military research establishment. If so, I guess I will never know. Apparently the US government thinks I'm not to be trusted.


\acknowledgements

I am indebted to many people for detailed discussions and collaborations on photonic quantum computing over the last few years. They include 
Hussain Anwar,
James Auger,
Sara Bartolucci,
Damien Bonneau,
Dan Browne,
Hugo Cable,
Jacques Carolan,
Mercedes Gimeno-Segovia,
Anthony Laing,
Jonathan Matthews, 
Gabriel Mendoza,
Sam Morley-Short,
Jeremy O'Brien,
Josh Silverstone,
Pete Shadbolt, 
Tom Stace,
and
Mark Thompson.

In particular I was dragged back into working seriously in this area by Gabriel, Jacques, Josh and Mercedes' initiation of the `QNIX' project, their creativity and enthusiasm coupled with strong technical ability is wonderful and inspiring. The work in Section \ref{sec:physicalarchitecture} on the physical layout is in collaboration with these four as well as Pete and Dan. Various parts of the work on the logical architecture in Section~\ref{sec:logicalarchitecture} are being done in collaboration with various combinations of Dan, James, Hussain, Mercedes, Pete, Sam, Sara and Tom. The work on multiplexing is being done in collaboration with Gabriel, Hugo, Jacques, Josh, Mercedes and Pete. The work on abstract loss tolerance in Section~\ref{sec:losstolerance} is being done with Mercedes and Dan.   
 
Finally, I am grateful to Earl Campbell and Joe O'Gorman for kindly sharing some of their results in advance of publication. 
\vspace{-0.3cm}

\bibliography{LOQCpapers.bib}
\bibliographystyle{ieeetr}

\appendix

\section{Sociological musings}

Photonics benefits from ``quick and dirty'' sources of small amounts of entanglement (via spontaneous parametric downconversion) and so most of the initial demonstrations of quantum information processing were performed in an inherently unscalable fashion. Implicitly, and sometimes explicitly, I hear the view expressed that rather than engineer a difficult new technology (good non-random photon sources) workers in photonics just added a beamsplitter here or there, occasionally added another non-deterministic source, then waited a bit longer to collect enough data to creatively re-package into a wide variety of Nature and Science papers. This, the argument goes, has led to over-resourcing of photonic approaches relative to its potential. 


Inevitably different routes to building a quantum computer will compete for resources. That time should not be now. Given the unbelievable potential for this technology and its expected related spinoffs in communication, high precision measurement as well as things unforeseen, we should only be limited by the availability of skilled researchers, not our ability to pay them or give them suitable equipment. Every route to quantum computing that is being pursued seriously is pushing us to deeper understanding of the physics of the systems involved, as well as pushing engineering boundaries. As a community we now have a wide variety of very different types of physical systems over which we are gaining exquisite control at the scale of the individual constituents. For all of them there is real hope of generating large scale entanglement - the ultimate of all quantum weirdnesses - in the near future. 

The world has already been through a ``first quantum revolution'', where the much-less weird of the quantum weirdnesses (e.g. discrete energies, tunnelling, superposition, Bose-condensation) gave us corresponding devices (e.g. transistors, electron microscopes, atomic clocks, lasers) which in turn were crucial enablers of technologies even the thickest politician can see are worth 10's of trillions of any currency (e.g. computers, the GPS system, the internet). Much as very different systems are necessary to exploit these 1st-generation quantum technologies, it would be very surprising if a single type of physical system magically worked best for all 2nd-generation quantum technologies. Couple this with the fact that almost all serendipitous scientific discoveries come when we push a physical boundary (making stuff colder, cleaner, smaller etc than ever before), which is certainly what experimental quantum information science is doing, and the case to support all serious efforts at performing controlled quantum dynamics should be closed.            

Be that as it may, any individual scientist has finite resources, and so has to pick what to work on. The mundane, unspoken drivers of this choice are normally what specific physics we understand best given our background, and the fact you can only raise funding for stuff you are already acknowledged as an expert in. As such few people work with systems much different from those they did their early research in, despite us all knowing this was primarily an accident of geography. The spoken justification for our choice, however, is often of the form ``my choice is going to be the winner of the race'' (a fact we should blame partly on our individual human competitiveness and partly on funding agencies and their government overlords, who expect such trite from us). Workers in quantum photonics can, I think, say this with the same level of conviction as those in any other area.


\end{document}